\documentclass[a4paper,11pt]{article}
\usepackage{amssymb}
\usepackage{amsmath}
\usepackage[dvips]{graphicx}
\usepackage{pstricks}
\usepackage{pst-plot}
\usepackage{pst-node}

\newtheorem{theo}{Theorem}%[section]
\newtheorem{pro}{Proposition}
\newtheorem{rem}{Remark}

\title{Blow-up results for some\\ second order hyperbolic
inequalities\\ with a nonlinear term with respect to the velocity}
\author{M. Jazar and R. Kiwan\\Lebanese University and Universit\'e de Tours}
\begin{document}

\maketitle

\begin{center}
{\bf\small Abstract} \vspace{3mm}

\hspace{.05in}\parbox{4.5in} {\small We give sufficient conditions
on the initial data so that a semilinear wave inequality blows-up
in finite time. Our method is based on the study of an associated
second order differential inequality. The same method is applied
to some semilinear systems of mixed type.}
\end{center}



\section{Introduction}
Let $\Omega$ a bounded regular\footnote{The regularity needed here
is that ensuring the existence of a first eigenvalue with positive
associated eigenfunction for $L$.} domain of $\mathbb{R}^N$ and
let us denote respectively by $|\cdot|$ and $(\cdot,\cdot)$ the
natural norm and inner product of the space $H=L^2(\Omega)$. Let
$V\hookrightarrow H$ be a real Hilbert space that is continuously
and densely embedded into $H$. Denote by $\|\cdot\|$ and
$<\cdot,\cdot>_V$ respectively the norm and the inner product of
$V$ and by $L\in\mathcal{L}(V,V^*)$ the unique linear continuous
operator from $V$ to $V^*$, the dual of $V$, satisfying
$$<Lu,v>_{V^*,V}=<u,v>_V\qquad (u,v)\in V\times V.$$
Moreover we assume that the operator $L$ admits an eigenvalue
$\lambda$ (that we may assume to be positive without loss of
generality) with a positive associated eigenfunction $\varphi\in
V$.

Consider the semilinear evolution second order partial
differential inequality
\begin{equation}\label{eq1}\left\{\begin{array}{ll}
%u\in C\left([0,T);V\right)\cap C^1(\left[0,T);H\right)&\\
u''+L u \ge g(u')& \textrm{in }\mathcal{D}^*((0,T)\times\Omega),\\
u(0,x)=u_0(x)&\textrm{on }\Omega\\
u_t(0,x)=u_1(x)&\textrm{on }\Omega,\end{array}\right.
\end{equation}
on the phase space $(u_0,u_1)\in V\times H$.

The model problem is a semilinear wave equation with Dirichlet
boundary conditions, that is $V=H^1_0(\Omega)$, $L=-\Delta$, see
the book of S. Alinhac \cite{A} for a short view on semilinear
hyperbolic problems and the question of blow-up. The particular
case of equation with $g(x)=-x|x|^r$ is a dissipative semilinear
wave equation that has been extensively studied in the literature,
and many results on the existence of global solutions as well as
the decay estimates have been established. See for instance
\cite{Ca,C,HZ,S2}.\\ The case of equation with $g(x)=x|x|^r$ is
more complicated. For instance, a local existence-uniqueness of
solutions is only guaranteed for small $r$ and more regular
initial data. See for instance \cite{Ca, G} for equations with
nonlinearities on $u$ and $u_t$. In \cite{H}, A. Haraux showed
that this problem (with $g(x)=x|x|^x$) has no nontrivial global
and bounded solutions \cite[Proposition 1.2]{H}. Moreover, he
constructed examples of blowing-up solutions with arbitrarily
small initial data \cite[Proposition 1.4 and Corollary 1.5]{H}.
See also the work of H. Levin, S. R. Park and J. Serrin \cite{L}
and the references therein for a study of the long-time behavior
of solutions for equations with $f(u)-g(u_t)$ on the right hand
side of the equation on $\Omega=\mathbb{R}^N$.\\
In this paper we consider Problem (\ref{eq1}) where $g$ is a
positive convex function $C^1(\mathbb{R},\mathbb{R^+})$ satisfying
the growth condition
\begin{equation}\label{growth}g(x)\geq C
|x|^q\qquad \mbox{for all }x\in\mathbb{R},\end{equation} where
$C>0$ is a constant and $q>1$ if the dimension $N=1$ or 2 or
$1<q\le \frac {N+2}{N-2}$ if $N\ge 3$. We provide sufficient
conditions on the initial data $(u_0,u_1)$ so that the
solution of Problem (\ref{eq1}) blows-up in finite time.\\
The method used here is the so-called ``eigenfunction method''
which leads to the study of a second order ordinary differential
inequality that we make using invariant regions. This is the
purpose of the second section. In the third section we apply this
to the semilinear wave equation (\ref{eq1}). \textbf{S. Kaplan
\cite{Kaplan} was the first to use Fourier coefficient's method in
this setting. Later, R. Glassey \cite{Glassey} applied this method
to nonlinear wave equations. In particular he treated an equation
similar to (\ref{eq1}) in the whole space $\mathbb{R}^N$. H.
Levine \cite[Section 5]{Levine} considered equation similar to
(\ref{eq1}) on bounded domain with Dirichlet boundary conditions.
Under some hypotheses on the nonlinearity $g$, the eigenvalue
$\lambda$ and the initial data, Levine gives a necessary condition
for finite time blow-up. Our assumptions on the nonlinearity are
different, in particular $g$ does not need to be nondecreasing.
Moreover, the method we use for the study of the ODI is different
from that given in \cite{Levine}. However, the model case, the
case of equality with $g(s)=C|s|^q$, is covered by Levine's work
(see \cite[Section 5]{Levine}), but our assumptions on the initial
data are weaker. In the third Section (Remark \ref{comparison}) we
give a comparison between Levine's result and ours.}\\ In the
fourth section we apply our result to semilinear wave systems of
the form:
\begin{equation}\label{eqS}\left\{\begin{array}{ll}
u_{tt}-\Delta u&\ge |v_t|^p\\
v_{tt}-\Delta v&\ge |u_t|^q,
\end{array}\right.\end{equation}
by means of an extension of results of Section 2 to an associated
system of second order ordinary differential inequalities. In the
last section we obtain similar results for systems of mixed type:
hyperbolic-elliptic and hyperbolic-parabolic systems (see for
instance the work of Pohozaev and V\'eron \cite{PV}).

\noindent\textbf{Acknowledgment.} The authors thank the anonymous
referee of this paper for his report. He pointed out to us the
papers of R. Glassey \cite{Glassey} and H. Levine \cite{Levine}.

%%%%%%%%%%%%%%%%%%%%%%%%%%%%%%%%%%%%%%%
\section{Finite time blow-up for a second order ordinary
differential inequality}\label{section2} Consider the ODI:
\begin{equation}\label{ODI}\left\{\begin{array}{cl}
v''+av\geq bv'^q,& t\geq 0\\
v(0)=v_0,\qquad v'(0)=v_1,&
\end{array}\right.\end{equation} where $a,b$ are positive constants.

Notice that in \cite{BJS, S1} one can find a complete study of the
equation: $u''+a|u|^{p-1}u=b|u'|^{q-1}u'$ where $a,b>0$ and
$p,q>1$. In \cite{S11, S22}, Ph. Souplet proved the existence of
some global solutions to equation (\ref{ODI}).

\bigskip

Concerning the ODI (\ref{ODI}) we distinguish two cases: $q\le2$
and $q>2$
\begin{pro}\label{pro1} Let $q\leq 2$,
$\displaystyle \alpha
=\frac{2a}{bq}\left(\frac{4a}{b^2q}\right)^{\frac{2-q}{2q-2}}$
and $v_0,v_1\in\mathbb{R}$ satisfying
\begin{equation}\label{c2}\displaystyle\left|\begin{array}{ll} 0<v_1,\\ \\
av_0+\alpha<\frac b 2 v_1^q   &\textrm{if}\ \  v_1 <[\frac{4a}{b^2q}]^\frac 1{2q-2},\\ \\
av_0< \frac b 2 v_1^q & \textrm{if}\  \ v_1 \geq [\frac{4a}{b^2q}]^\frac 1{2q-2}.\\
\end{array}\right.\end{equation}
Then all solutions of (\ref{ODI}) blow-up in finite time.
Moreover,
\begin{equation}\label{rate}v'(t)\geq \left[v_1^{1-q}-\frac{q-1}{2}bt
\right]^{\frac{1}{1-q}}.\end{equation}
\end{pro}

\bigskip

The proof of the proposition is based on the idea that the
inequality $v''\ge bv'^q$ provide blowing up solution in finite
time $T<\frac{v_1^{1-q}}{b(q-1)}$. We look then for invariant
regions under the ODI (\ref{ODI}) in order to obtain $av^p\leq
b\varepsilon v'^q$.
\bigskip

\noindent\textbf{Proof} of Proposition \ref{pro1}. We show first
that the region $\mathcal{D}$ defined by (\ref{c2}) is invariant
under (\ref{ODI}). Denote by $F$ the function defined over
$\mathbb{R}$ by
$$F(x):=\left\{\begin{array}{ll}
0&\mbox{for }x\le x_1:=-\frac\alpha a\\
\\
{\left[\frac2b(ax+\alpha)\right]}^{\frac 1q}&\mbox{for }
x_1\le x\le 0 \\% car ta region est plus petite
\\
{\left[\frac{4a}{b^2q}\right]}^\frac 1{2q-2}&\mbox{for }0\le x\le
x_2:=
\frac{b}{2a} \left[\frac{4a}{b^2q}\right]^\frac q{2q-2}\\
\\
{\left[\frac{2a}b x \right]}^{\frac 1q}&\mbox{for }x\ge x_2.
\end{array}\right.$$
The region $\mathcal{D}$ is then defined by $[y>F(x)]$.\\
Let $v$ be a solution of (\ref{ODI}) satisfying (\ref{c2}).
Setting $x=v$ and $y=v_t$, the ODI (\ref{ODI}) can be transformed
into the dynamical system:
\begin{equation}\label{sysODI}
\left\{\begin{array}{lcl}\displaystyle x'&=&P(x,y):=y,\\ y'&=&Q(x,y)\ge by^q -ax,\\
x(0)&=&v_0,\\ y(0)&=&v_1.
\end{array}\right.\end{equation}
In order to prove that the region $\mathcal{D}$ is invariant we
will show that the vector field defining the dynamical system is
``entering'' along the curve $y=F(x)$. This is clear on the
semi-axis $(y'O)$ and on the other segment. On the arc $x> x_2$,
$$\frac {y'} {x'}\geq \frac {by^q-ax}{y}
\geq\frac b 2 y^{q-1}\geq\frac{2a}{bqy^{q-1}}= F'(x).$$ Now, on
the arc $x_1< x< 0$ we have
$$\frac{y'}{x'}\geq \frac \alpha y\ge F'(x).$$
Therefore for all $t>0$ we have  $\displaystyle\frac b 2 v_t^q\geq
av$, and then $v''>\frac 12 bv'^q$. Integrating we get
(\ref{rate}). ${}\hfill\square$

\bigskip
The following proposition gives a similar result for the case
$q>2$. we discuss later the difference between the corresponding
invariant regions in these two cases.

\begin{pro}\label{pro1*} Let $q>2$ and $v_0,v_1\in\mathbb{R}$ satisfying
\begin{equation}\label{c2*a}\displaystyle\left|\begin{array}{l} 0<v_1,\\
bqv_1^{q-2}(bv_1^q-av_0)-a  >0 .\\
\end{array}\right.\end{equation}
Then all solutions of (\ref{ODI}) blow-up in finite time.
Moreover,
\begin{equation}\label{rate1}v'(t)\geq \left[v_1^{1-q}-(q-1)(1-\varepsilon)bt
\right]^{\frac{1}{1-q}}\end{equation} where $\varepsilon\in(0,1)$
is a constant satisfying: \begin{equation}\label{c2*b}\varepsilon
bqv_1^{q-2}(\varepsilon bv_1^q- av_0) -a\geq 0.\end{equation}
\end{pro}
\noindent\textbf{Proof.} First note that, using (\ref{c2*a}) and
taking $$f(\varepsilon)=\varepsilon bqv_1^{q-2}(\varepsilon
bv_1^q- av_0) -a,$$ there exists at least an $\varepsilon\in
(0,1)$ satisfying $f(\varepsilon)>0$.\\
Setting $A=\varepsilon b v_1^q -av_0$, we have $v_1^{q-2}\geq\frac
a{\varepsilon bqA}$. Denote by $\mathcal{D}'$ the domain
$[y>F_2(x)]$, where $F_2(x):={[\frac 1{\varepsilon b(ax +A)}]}^{\frac 1q}$.
Since $y'$ remains positive, as long as $(x,y)$ belongs to
$\mathcal{D}'$, then for all $t>0$, $\left(y(t)\right)^{q-2}\geq\frac a {\varepsilon
bqA}$. Following the same arguments as in the last proposition, we
have, along the curve $y=F_2(x)$,
\begin{eqnarray*}
\frac {y'} {x'}\geq \frac{by^q -ax}{y}\geq\frac{\varepsilon by^q
-ax}{y} =\frac A y \geq\frac a {\varepsilon bqy^{q-1}}=
F_2'(x).\end{eqnarray*} ${}\hfill\square$

\begin{rem}\label{admissible}\textbf{(On the admissible initial data region)}\\
In the following figures we draw the curve limiting the region for
two values of $q$. The admissible region in each case is colored
in gray.\\
\begin{center}
\pspicture(-3,0)(3,2.7)
\pscustom[fillstyle=solid,fillcolor=lightgray,linecolor=white]{
\psline(-3,0)(-2.17,0) \psplot{-2.17}{0}{x 2.17 add .66 exp}
\psline(0,1.66)(.76,1.66) \psplot{.76}{3}{x 1.4 add .66 exp}
\psline(-3,2.65)
 }
 \psline(-3,0)(-2.17,0) \psplot{-2.17}{0}{x 2.17
add .66 exp} \psline(0,1.66)(.76,1.66) \psplot{.76}{3}{x 1.4 add
.66 exp}

\psline{->}(-3,0)(3,0)\psline{->}(0,0)(0,2.65)
\rput(2.7,.2){$v_0$}\rput(-.3,2.45){$v_1$}
\endpspicture
\hskip 1cm \pspicture(-2,0)(3,2.7)
\pscustom[fillstyle=solid,fillcolor=lightgray,linecolor=white]{
\psline(-2,2.7)(-2,.01)
\parametricplot{.04}{1.616}{t 2.5 exp .4 t -.5 exp mul sub t}
\psline(3,2.7) \psline(-2,2.7)}
\parametricplot{.04}{1.616}{t 2.5 exp .4 t -.5 exp mul sub t}
\psline{->}(-2,0)(3,0)\psline{->}(0,0)(0,2.7)
\rput(2.7,.2){$v_0$}\rput(-.3,2.5){$v_1$}
\endpspicture
\end{center}

{\small $a=1$, $b=2$ and $q=1.5$. \hskip 1.7in $a=b=1$ and
$q=2.5$.}

\end{rem}

\bigskip

\begin{rem}\label{rem2}
The same method works for the differential inequality
$$v''+av^p\geq bv'^q,$$
where $p\le 1$.
\end{rem}

%%%%%%%%%%%%%%%%%%%%%%%%%%%%%%%%%%%%
\section{Blow-up criteria for a semilinear wave
inequality}\label{section3} In this section we assume that we have
a Sobolev injection type theorem: $V\hookrightarrow
L^{2^*}(\Omega)$, with $2^*=\frac{2N}{N-2}$. Consider the problem
\begin{equation}\label{eq2}\left\{\begin{array}{ll}
u\in C\left([0,T);V\right)\cap C^1(\left[0,T);H\right)&
%\capC^2((0,T); V^*),
\\
u''+L u \ge g(u')& \textrm{in }\mathcal{D}^*((0,T)\times\Omega),\\
u(0,x)=u_0(x)&\textrm{on }\Omega\\
u_t(0,x)=u_1(x)&\textrm{on }\Omega,\end{array}\right.
\end{equation}
where $g$ is a positive convex function
$C^1(\mathbb{R},\mathbb{R^+})$ satisfying the growth condition
(\ref{growth}), $q>1$ if the dimension $N=1$ or 2 or $1<q\le \frac
{N+2}{N-2}$ if $N\ge 3$ and $(u_0,u_1)\in V\times H$. Indeed,
local existence is guaranteed under hypotheses of this type on
both the power $q$ and the smoothness of the initial data, see for
instance \cite{CH,H}. Under such hypotheses, local existence can
be easily obtained by classical fixed point argument and abstract
semi-group theory.

Let $\lambda$ be the first positive eigenvalue of the operator $L$
on $H$, and $\varphi$ a nonnegative associated eigenfunction
satisfying $\int_{\Omega}\varphi=1$. Denote also by
$v_i:=\int_\Omega u_i\varphi$,  $i=0,1$. As for the ODI
(\ref{ODI}), we distinguish 2 cases depending on the value of $q$.

\begin{theo}\label{th1}
Let $(u_0,u_1)\in V\times H$ such that $v_0$ and $v_1$ satisfy:
\begin{displaymath}\begin{array}{ll}

 \textrm{if}\quad q\leq 2&\left|\begin{array}{l} 0<v_1,\\ \\
\begin{array}{ll}
\lambda v_0+\alpha<\frac C 2 v_1^q &
 \textrm{if}\ v_1 <(\frac{4\lambda}{C^2q})^\frac 1{2q-2},\\ \\
\lambda v_0\leq \frac C 2 v_1^q
&\textrm{if}\ v_1 \geq (\frac{4\lambda}{C^2q})^\frac 1{2q-2}.\\
\end{array}\end{array}\right.\\ \\
\displaystyle \textrm{and if}\quad q> 2&\left|\begin{array}{l}
0<v_1,\\ \\
Cqv_1^{q-2}(Cv_1^q-\lambda v_0)-\lambda  >0 .\\
\end{array}\right.\end{array}\end{displaymath}
where $\alpha:=\frac{2\lambda}{Cq} \left(\frac{4 \lambda}
{C^2q}\right)^\frac {2-q} {2q-2}$. Then every solution of
(\ref{eq2}) blows-up in finite time
$$ T_{max}\le T^*:=\frac{v_1^{1-q}}{(1-\varepsilon)C(q-1)},$$
where $\varepsilon:=\frac 1 2$ if $q\leq 2$, and if $q>2$ ,
$\varepsilon$ satisfy $\varepsilon Cqv_1^{q-2}(\epsilon Cv_1^q-
\lambda v_0) -\lambda\geq 0$. Moreover,
$$\displaystyle\|u_t(t)\|_{L^1(\Omega)}\ge
\|\varphi\|_{L^\infty(\Omega)}^{-1}
\left[v_1^{1-q}-(q-1)(1-\varepsilon)Ct \right]^{\frac{1}{1-q}}.$$
\end{theo}
\noindent\textbf{Proof.} Set $v(t):=\int_\Omega u(t)\varphi\,dx$.
By the elliptic regularity theory $\varphi\in V$ and then $v$ is
twice differentiable with $v'=\int u_t\varphi$ and
$v''=<u_{tt},\varphi>_{V^*,V}$ that we write also as $\int_\Omega
u_{tt}\varphi\,dx$.

Multiply the equation (\ref{eq2}) by $\varphi\ge 0$ and then
integrate over $\Omega$, we get after integrating by parts
\begin{eqnarray*}
\int_\Omega u_{tt}\varphi\,dx-\int_{\Omega}L u\varphi\,dx&\ge&
C\int_\Omega|u_t|^q\varphi\,dx\\
v''(t)+\lambda v(t)&\ge& C|v'(t)|^q,
\end{eqnarray*}
where we used Jensen's lemma. Using Propositions \ref{pro1} and \ref{pro1*} we
deduce that $v$ blows-up before the time $T^*$ and
\begin{eqnarray*}\|\varphi\|_{L^\infty(\Omega)}\|u_t(t)\|_{L^1(\Omega)}
&\ge& \left|\int_\Omega u_t(t)\varphi \,dx\right|=|v'(t)|\\ &\ge&
\left[v_1^{1-q}-C(q-1)(1-\epsilon)t \right]^{\frac{-1}{q-1}}.
\end{eqnarray*}
${}\hfill\square$

\begin{rem}\label{comparison} \textbf{(Comparison with the
results of Levine \cite{Levine})}\\
In \cite[Theorem 5.1]{Levine}, Levine shows a blow-up result for a
class of equations. Levine's result can be easily applied to
equation (\ref{eq2}) with $g(s)=C|s|^q$. However, the assumptions
he needs on the initial data are not the same. Indeed, by
\cite[Hypothesis (5.5)]{Levine}, $v_0$ and $v_1$ satisfy
$$v_1>v_0>s_0$$
for some $s_0$ ($s_0$ should be in this case greater than
${\left(\frac{\lambda+1}C\right)^{\frac1{q-1}}}$). This gives a
region of the form\\
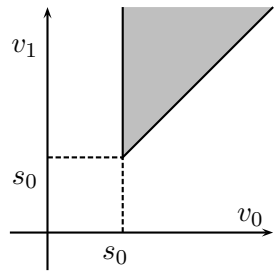
\begin{center}
\pspicture(-.5,-.5)(3,3)
\pscustom[fillstyle=solid,fillcolor=lightgray]
{\psline(1,3)(1,1)(3,3)}

\psline[linestyle=dashed,dash=2pt 1pt](0,1)(1,1)(1,0)
\rput(-.3,.7){$s_0$} \rput(.9,-.3){$s_0$}

\psline{->}(-.5,0)(3,0)\psline{->}(0,-.5)(0,3)
\rput(2.7,.2){$v_0$}\rput(-.3,2.45){$v_1$}
\endpspicture
\end{center}

\noindent to be compared with the regions of Remark
\ref{admissible}.

\end{rem}

\begin{rem}
According to Remark \ref{rem2}, the same method could be
generalized to inequalities of the form ($p\le 1$)
$$u_{tt}-\Delta u^p\ge g(u_t).$$
\end{rem}

%%%%%%%%%%%%%%%%%%%%%%%%%%%%%%%%%%%%%%%%%%%
\section{Application to systems of semilinear wave inequalities}
In this section we show how to extend the same method to the
following system
\begin{equation}\label{sysonde}
\left\{\begin{array}{ll} u,v\in C\left([0,T);V\right)\cap
C^1(\left[0,T);H\right)& %\cap C^2((0,T); V^*),
\\
\left.\begin{array}{l}u''+ Lu \ge |v'|^p\\
v''+L v \ge |u'|^q\end{array}\right\}&\textrm{in }\mathcal{D}^*((0,T)\times\Omega),\\
u(0,x)=u_0(x),\quad u_t(0,x)=u_1(x)&\textrm{on }\Omega\\
v(0,x)=v_0(x),\quad v_t(0,x)=v_1(x)&\textrm{on }\Omega,
\end{array}\right.
\end{equation}
where $(u_0,u_1)$ and $(v_0,v_1)$ are in $V\times H$ and the
powers $p$ and $q$ are greater than 1 if the dimension $N=1$ or 2
or in $\left(1, \frac {N+2}{N-2}\right]$ if $N\ge 3$. For
simplicity we choose a simple form of the non linearity, although
a general form is possible. For this, we introduce the following
system of ODI:
\begin{equation}\label{eq14}\left\{\begin{array}
{cl}U''+aU\geq V'^p&t\geq 0\\
V''+aV\geq U'^q&t\geq 0\\
U(0)=U_0,\qquad U'(0)=U_1&\\
V(0)=V_0,\qquad V'(0)=V_1.&
\end{array}\right.\end{equation}
The method of Section \ref{section2} could be easily extended to
the previous system as follows:
\begin{pro}\label{pro2}
Let $1<p\le q$, $U_0,V_0,U_1$ and $V_1$ satisfying
\begin{equation}\displaystyle\label{c4}\left|\begin{array}{rcl}
U_1,V_1>1,&&U_0V_0\ge U_1V_1,\\
U_1^{q} &\geq& [a+\textstyle{\frac1p}]V_0,\\ \\ V_1^{p} &\geq&
[a+\textstyle{\frac1p}]U_0.
\end{array}\right.\end{equation}
Then every solution $(U,V)$ of (\ref{eq14}) blows-up in finite
time. Moreover we have
$$U'(t)+V'(t)\geq \left[(U_1+V_1)^{1-p}-\frac{p-1}{1+ap}2^{1-p}t
\right]^{\frac{1}{1-p}}.$$
\end{pro}
\textbf{Proof.} Setting $x=U,\ y=V',\ z=V\ and\ t=U'$,  then the
system (\ref{eq14}) is transformed into the 4-d dynamical system
\begin{equation*}\left\{\begin{array}{ll}\displaystyle
x'=t,&x(0)=U_0,\\
t'\geq y^p -ax, &t(0)=U_1,\\
z'=y, &z(0)=V_0,\\
y'\geq t^q -az,&y(0)=V_1.
\end{array}\right.\end{equation*}
Set $\alpha:=1+\frac 1{ap}$ and $f_p(x):=H(x)(\alpha ax)^{\frac
1p}$, where $H$ is the Heaviside function. Denote by
$\mathcal{D}_1:=[y\ge f_p(x)]$ and $\mathcal{D}_2:=[t\ge f_q(z)]$.
The hypothesis (\ref{c4}) can then be read as
$(U_0,U_1)\in\mathcal{D}_1$ and $(V_0,V_1)\in\mathcal{D}_2$.\\
First notice that as long as $(x,y,z,t)$ remains in
$\mathcal{D}_1\times\mathcal{D}_2$, $U''$ and $V''$ remain
positive, so for all $t>0$, $t^{q-1}y^{p-1}\geq U_1^{q-1}V_1{p-1}
=\frac{U_1^qV_1^p}{U_1V_1}\ge \alpha^2 a^2\frac{U_0V_0}{U_1V_1}\ge \alpha^2a^2$.\\
In order to show that $(x,y,z,t)$ could not exit the domain
$\mathcal{D}_1\times\mathcal{D}_2$ let's examine the vector field
along the boundary. On $[t=f_q(z)]$ we have
\begin{eqnarray*}
\frac {y'} {x'}& \geq&\frac{t^q-az}t=\frac {(\alpha-1)az}t=\frac{\alpha-1}\alpha t^{q-1}\\
&\geq&\frac{\alpha-1}\alpha \frac{\alpha^2a^2}{y^{p-1}}=
ap(\alpha-1)f_p'(x)= f_p'(x).
\end{eqnarray*}
A similar calculation holds for the other boundary.\\
Finally, setting $\beta:=\frac{\alpha-1}\alpha$, we have:
\begin{equation*}\left\{\begin{array}{ll}\displaystyle
U''&\geq \beta V'^p,\\
V''&\geq \beta U'^q.
\end{array}\right.\end{equation*}
Adding these two equations, using the fact that $U'\ge 1$ and
$V'\ge 1$ for all $t>0$, we get:
\begin{equation}\frac 1{2\beta}W''\ge \frac 1 2 \left(V'^p + U'^q\right)
\geq\frac 1 2 \left(V'^p + U'^p\right)\geq
2^{-p}W'^p,\end{equation} where $W:=U+V$. Integrating we obtain:
$$W'(t)\geq \left(W_1^{1-p}-(p-1) \beta2^{1-p}t\right)^{\frac{1}{1-p}}.$$
${}\hfill\square$
\bigskip

In order to apply the last result to the system (\ref{sysonde}),
denote by $U_i:=\int_\Omega u_i\varphi$ and $g_i:=\int_\Omega
v_i\varphi$ , $i=0,1$. Using the same method as in the proof of
Theorem \ref{th1} and applying the Proposition \ref{pro2} we get
directly:

\begin{theo}\label{th2}
Let $1<p\le q$, $(u_0,u_1)$ and $(v_0,v_1)$ in $V\times H$ such
that $U_0,V_0,U_1$ and $V_1$ satisfy the hypothesis
\begin{equation}\displaystyle\label{c5}\left|\begin{array}{rcl}
U_1,V_1>1 &&U_0V_0\ge U_1V_1,\\
U_1^{q} &\geq& [\lambda+\textstyle{\frac1p}]V_0,\\ \\
V_1^{p}
&\geq& [\lambda+\textstyle{\frac1p}]U_0.
\end{array}\right.\end{equation}
Then every solution of (\ref{sysonde}) blows-up in finite time.
Moreover, we have
$$\|u_t+v_t\|_{L^1(\Omega)}\geq \|\varphi\|_{L^\infty(\Omega)}^{-1}
\left[(U_1+V_1)^{1-p}-\frac{p-1}{1+\lambda p}2^{1-p}t
\right]^{\frac{1}{1-p}}.$$
\end{theo}

%%%%%%%%%%%%%%%%%%%%%%%%%%%%%%%%%%%%%%%%%%5\angle
\section{Application to systems of mixed types}
In this section we show how to apply the result of section
\ref{section3} to some hyperbolic-elliptic and
hyperbolic-parabolic systems.

\subsection{Hyperbolic-elliptic system}
Consider the following system
\begin{equation}\label{chemo}
\left\{\begin{array}{ll} u,v\in C\left([0,T);H^1_0(\Omega)
\right)\cap C^1(\left[0,T);L^2(\Omega)\right)& %\cap C^2((0,T); V^*),
\\
\left.\begin{array}{l}u_{tt}-\Delta u \ge |v_t|^q\\
-\Delta v = u \end{array}\right\}&\textrm{in }\mathcal{D}^*((0,T)\times\Omega),\\
u(0,x)=u_0(x),\quad u_t(0,x)=u_1(x)&\textrm{on }\Omega\\
v(0,x)=v_0(x)&\textrm{on }\Omega,
\end{array}\right.
\end{equation}
where $q>1$.
%Such systems could describe chemotaxis mouvement in microscopic scale...see [?]...

%We give the following blow-up criteria

\begin{theo}
Let $u_0,v_0\in H^1_0(\Omega)$ and $u_1\in L^2(\Omega)$ such that
$U_0:=\int_\Omega u_0\varphi\,dx$ and $U_1:=\int_\Omega
u_1\varphi\,dx$ satisfy:
\begin{displaymath}\begin{array}{rll}
\textrm{if}\ & q< 2& \displaystyle\left|\begin{array}{l} 0<U_1,
\\ \\
\begin{array}{lcl}\lambda U_0+\alpha<\frac {\lambda^{-q}} 2 U_1^q
&\textrm{if}& U_1 <(\frac{4\lambda^{2q+1}}{q})^\frac 1{2q-2},\\ \\
\lambda U_0\leq \frac {\lambda^{-q}} 2 U_1^q & \textrm{if}& U_1
\geq (\frac{4\lambda^{2q+1}}{q})^\frac 1{2q-2}.\\
\end{array}\end{array}\right.
\\ \\
\textrm{and if}& q> 2& \displaystyle\left|\begin{array}{l}
0<U_1,\\ \\
q\lambda^{-q}U_1^{q-2}(\lambda^{-q}U_1^q-\lambda U_0)-\lambda  >0,\\
\end{array}\right.
\end{array}\end{displaymath}
where $\alpha:=\frac{2\lambda^{1+q}}q
\left(\frac{4\lambda^{2q+1}}q\right)^{\frac{2-q}{2q-2}}$.
 Then every solution of
(\ref{chemo}) blows-up in finite time
$$ T_{max}\le T^*:=\frac{U_1^{1-q}\lambda^q}{(1-\varepsilon)(q-1)},$$
where $\varepsilon:=\frac 1 2$ if $q\leq 2$, and if $q>2$ ,
$\varepsilon$ satisfy $\varepsilon Cqv_1^{q-2}(\epsilon Cv_1^q-
\lambda v_0) -\lambda\geq 0$. Moreover,
$$\displaystyle\|u_t(t)\|_{L^1(\Omega)}\ge
\|\varphi\|_{L^\infty(\Omega)}^{-1}
\left[U_1^{1-q}-(q-1)(1-\varepsilon)\lambda^{-q}t \right]^{\frac{1}{1-q}}.$$
\end{theo}

\noindent\textbf{Proof.} Denote by $U:=\int_\Omega u\varphi\, dx$
and $V:=\int_\Omega v\varphi\,dx$. Differentiate the second
equation of (\ref{chemo}), multiply by $\varphi$, then integrate
over $\Omega$ we get
\begin{equation}\label{eq5}\lambda V'(t)=U'(t).\end{equation}
Multiplying the first equation of (\ref{chemo}) by $\varphi$ and
then integrating over $\Omega$ we get, after using (\ref{eq5}),
$$U''(t)+\lambda U(t)\ge \lambda^{-q}|U'(t)|^q.$$
We conclude applying the propositions \ref{pro1} and \ref{pro1*}
with $a=\lambda$ and $b=\lambda^{-q}$. $\hfill\square$

\subsection{Hyperbolic-parabolic system}

Consider the following system
\begin{equation}\label{chemo1}
\left\{\begin{array}{ll} u,v\in C\left([0,T);H^1_0(\Omega)
\right)\cap C^1(\left[0,T);L^2(\Omega)\right)& %\cap C^2((0,T); V^*),
\\
\left.\begin{array}{l}u_{tt}-\Delta u \ge |v_t|^q\\
(u-v)_t-\Delta (u-v)^m \le\beta (u-v)^p
\end{array}\right\}&\textrm{in }\mathcal{D}^*((0,T)\times\Omega),\\
u-v\ge 0&\textrm{on }[0,T)\times\Omega,\\
u(0,x)=u_0(x),\quad u_t(0,x)=u_1(x)&\textrm{on }\Omega\\
v(0,x)=v_0(x)&\textrm{on }\Omega,
\end{array}\right.
\end{equation}
where $q>1$, $m\ge 1\ge p$ and $\beta\in\mathbb{R}$.\\
%Such systems could ...

\begin{theo}
Let $u_0,v_0\in H^1_0(\Omega)$ and $u_1\in L^2(\Omega)$. Set
$U_0:=\int_\Omega u_0\varphi\,dx$, $V_0:=\int_\Omega
v_0\varphi\,dx$ and $U_1:=\int_\Omega u_1\varphi\,dx$. Assume that
$\beta\le 0$  or $\quad[ \beta\le \lambda \mbox{ and }U_0-V_0\ge 1
]$, $u_0-v_0\ge 0$ and
\begin{equation}\label{hyp}\begin{array}{rcl}
\textrm{if}\  &q< 2& \displaystyle\left|\begin{array}{l}
0<U_1,\\
\begin{array}{lcl}
\lambda U_0+\alpha<\frac 1 2 U_1^q
&\textrm{if}& U_1 <(\frac{4\lambda}{q})^\frac 1{2q-2},\\
\lambda U_0\leq \frac 1 2 U_1^q &\textrm{if}&U_1
\geq (\frac{4\lambda}{q})^\frac 1{2q-2}.\\
\end{array}\end{array}\right.\\ \\
\textrm{and if}& q> 2& \displaystyle\left|\begin{array}{l}
0<U_1,\\
qU_1^{q-2}(U_1^q-\lambda v_0)-\lambda  >0,\\
\end{array}\right. \end{array}\end{equation}
where $\alpha =\frac{2\lambda}q\left(\frac{4a}q\right)^{\frac{2-q}{2q-2}}$.
\\ Then every solution of (\ref{chemo1}) blows-up in
finite time
$$ T_{max}\le T^*:=\frac{U_1^{1-q}}{(1-\varepsilon)(q-1)},$$
where $\varepsilon:=\frac 1 2$ if $q\leq 2$, and if $q>2$ ,
$\varepsilon$ satisfy $\varepsilon qv_1^{q-2}(\varepsilon v_1^q-
\lambda v_0) -\lambda\geq 0$. Moreover,
\begin{eqnarray*}\|v_t(t)\|_{L^1(\Omega)}&\ge& \|u_t(t)\|_{L^1(\Omega)}\\&\ge&
\|\varphi\|_{L^\infty(\Omega)}^{-1}
\left[U_1^{1-q}-(q-1)(1-\varepsilon)t \right]^{\frac{1}{1-q}}.\end{eqnarray*}
\end{theo}

\noindent\textbf{Proof.}
Denote by $U:=\int_\Omega u\varphi\, dx$ and $V:=\int_\Omega v\varphi\,dx$.
Multiplying the second equation of (\ref{chemo1}) by $\varphi$ and
then integrating over $\Omega$ we get after using Jensen's lemma
\begin{eqnarray}\label{eq6}
(U-V)'&\le& \int_\Omega\left[\beta (u-v)^p -\lambda
(u-v)^m\right]\varphi\\&\le& \beta (U-V)^p-\lambda(U-V)^m.
\end{eqnarray}
By the hypothesis on $\beta$ and using the diffusion property,
$U'\le V'$. Set
$$T:=\max\left\{0\le t<T_{max}\mbox{ s.t. }U'(s)\ge 0\quad\forall
s\in[0,t]\right\}.$$ Since $U_1>0$, one has $T>0$. For all
$t\in[0,T)$ we have $0\le U'\le V'$ and, hence, $0\le U'^q\le
V'^q$. Multiplying the first equation of (\ref{chemo1}) by
$\varphi$ then integrating over $\Omega$ we get
\begin{equation}\label{eq7}U''+\lambda U\ge
U'^q\end{equation} for all $t\in [0,T)$. In order to apply
Propositions \ref{pro1} and \ref{pro1*} we need to prove that $T$
is large enough. Indeed, by the hypothesis (\ref{hyp}), $(U,U')$
remains in the invariant region defined by (\ref{hyp}) (see the
proof of Propositions \ref{pro1} and \ref{pro1*}). Thus $U'(t)>0$
for all $t$ and we obtain the result by applying Propositions
\ref{pro1} and \ref{pro1*} since (\ref{eq7}) holds for all $t$.
$\hfill\square$

\newpage

%%%%%%%%%%%%%%%%%%%%%%%%%%%%%%%%%%%%%%%%
%%%%%%%%%%%%%%%%%%%%%%%%%%%%%%%%%%%%%%%%

\vskip 1cm

\begin{center}
\begin{tabular}{cc}
M. Jazar&{}\hskip 2cm{}  R. Kiwan\\
Lebanese University&{}\hskip 2cm{}  Universit\'e de Tours\\
Mathematics Department&{}\hskip 2cm{} Laboratoire de Math\'ematiques\\
P.O.Box 155-012&{}\hskip 2cm{}  et Physique Th\'eorique\\
Beirut Lebanon&{}\hskip 2cm{} UMR 6083 du CNRS\\
mjazar@ul.edu.lb&{}\hskip 2cm{} Parc de Grandmont\\
&{}\hskip 2cm{} 37200 Tours France\\
&{}\hskip 2cm{} kiwan@lmpt.univ-tours.fr
\end{tabular}
\end{center}
\end{document}